\documentclass[times,3p,number,pdftex]{elsarticle}

\usepackage{amsmath} 
\usepackage{amsfonts}
\usepackage{graphicx} 
\usepackage{color}
\usepackage{subfigure}
\usepackage[latin2]{inputenc}
\usepackage{textcomp} 
\usepackage{multirow}
\usepackage{rotating}
\usepackage{units} 

\usepackage{floatpag}
\rotfloatpagestyle{empty}

\journal{Physica A}

\usepackage[normalem]{ulem}

\begin{document}

\begin{frontmatter}

\title{Impact of the tick-size on financial returns and correlations}

\author{Michael C. M\"unnix\corref{cor1}}
\ead{michael@muennix.com}
\author{Rudi Sch\"afer \corref{}}
\author{Thomas Guhr \corref{}}
\cortext[cor1]{Corresponding author. Tel.: +49 203 379 4727; Fax: +49 203 379 4732.}

\address{Fakult\"at f\"ur Physik, Universit\"at Duisburg-Essen, Germany}

\begin{abstract}
We demonstrate that the lowest possible price change (tick-size) has a large impact on the structure of financial return distributions. It induces a microstructure as well as it can alter the tail behavior. On small return intervals, the tick-size can distort the calculation of correlations. This especially occurs on small return intervals and thus contributes to the decay of the correlation coefficient towards smaller return intervals (Epps effect).
We study this behavior within a model and identify the effect in market data. Furthermore, we present a method to compensate this purely statistical error.
\end{abstract}


\end{frontmatter}

\section{Introduction}
The lowest possible price change of a financial security, the so called \emph{tick-size} or \emph{minimum tick}, plays an important role in quantitative finance. All raw price information is discretized by the tick-size. Historically, the tick-size of most securities has been consecutively reduced resulting in tick-sizes of \nicefrac{1}{100th}. This process is often referred to as decimalization. One reason for it was to aim at an enhanced market efficiency. In principle, small tick-sizes allow for a faster clearing of market arbritage. Nonetheless, it is controversial whether a smaller tick-size generally improves the market quality \cite{beaulieu04, huang01, declerck02, chung01}, e.g., in view of the fact that a larger tick-size ensures liquidity \cite{harris91}. Furthermore, a recent study indicates that in some cases only a fraction of the theoretically possible prices are used. Hence, prices cluster at certain multiples of the tick-size resulting in an effective tick-size \cite{onnela09}.

However, a large tick-size can lead to erroneous data in financial indices due to rounding errors \cite{kozicki09}.  The actual tick-size  for stocks is typically \$0.01. This is the case for instance on the New York Stock Exchange (NYSE) and the National Association of Securities Dealers Automated Quotations (NASDAQ). However, some securities such as U.S. Government Securities are still quoted in \nicefrac{1}{32nds} of a dollar.

The tick-size certainly affects many fields in quantitative finance. In this study we want to focus on its impact on two of the most important observables: relative price changes (\emph{financial returns}) and financial correlations. These elementary values are of particular importance for many applications, for example portfolio optimization \cite{markowitz52, doran09} and risk management \cite{b_bouchaud00}.

The article is organized as follows. In section \ref{s:returns}, we will study the influence of the tick-size on the microstructure of financial return distributions.
The impact of the tick-size on the calculation of financial correlations will be discussed in section \ref{s:corr}. The decay of the correlation coefficients towards small return intervals is of particular interest. This behavior is commonly referred to as the Epps effect \cite{epps79,reno03}. 
The identified mechanism is solely caused by the discrete tick-size and therefore represents a statistical effect. Hence, we are able to develop a method for compensating this distortion. We summarize the results in section \ref{s:conclusion}.

\section{Financial returns}
\label{s:returns}
Observations on financial data on very small time scales are usually referred to as market microstructure \cite{b_voit01micro}.
In this study, we will first investigate the influence of the tick-size on the shape and the microstructure of the financial return distribution. For this purpose, we decompose the set of returns according to the absolute price changes and disclose its microstructure. 

Subsequently we will demonstrate that this microstructure can alter the tail behavior of the return distribution compared to the underlying price change distribution.
Accordingly, we will disclose a relation between the tail behavior of each microstructure return distribution and the overall return distribution.

\subsection{Return microstructure}
\label{s:returnmicro}
A \emph{financial return} describes the relative price change of a security between two points in time. The arithmetic return is defined as
\begin{equation}
\label{eq:returndef}
r_{\Delta t}(t)=\frac{ \Delta S_{\Delta t}(t) }{S(t)}\ , 
\end{equation}
where $S(t)$ denotes the price at time $t$ and 
\begin{equation}
\Delta S_{\Delta t}(t) = S(t+\Delta t) - S(t)
\label{eq:DeltaS}
\end{equation}
is the (absolute) price change within the interval $[t,t+\Delta t]$.

\begin{figure}[h!]
\begin{center}
   \subfigure[]{
   \includegraphics[width=0.48\textwidth]{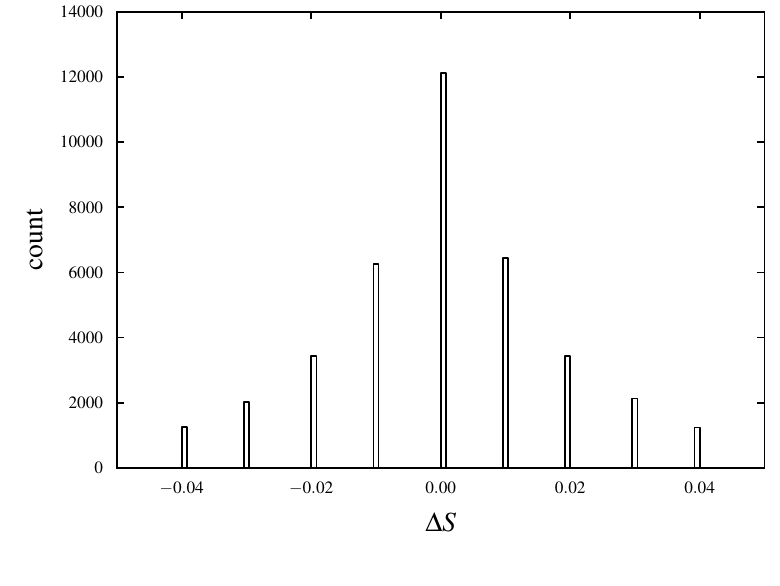}
     \label{fig:dS}
  }
\subfigure[]{
  \includegraphics[width=0.48\textwidth]{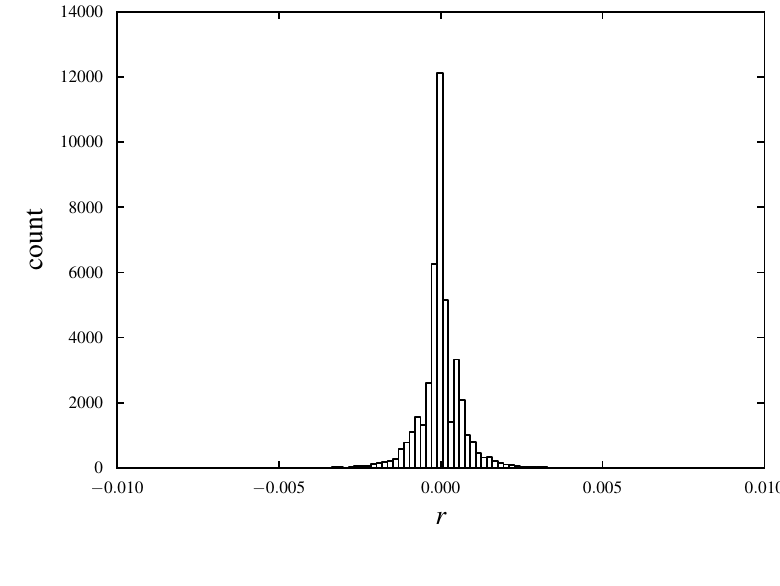}
  \label{fig:r}
  }
\\
 \subfigure[]{
   \includegraphics[width=0.48\textwidth]{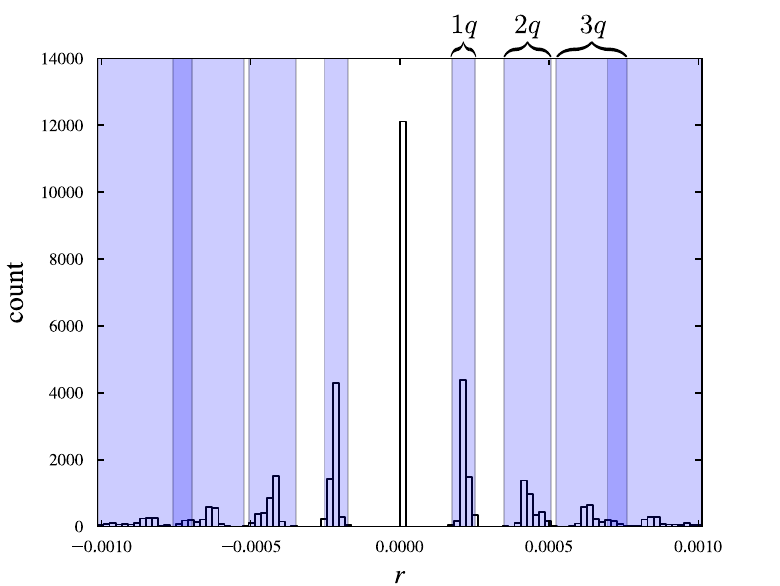}
     \label{fig:rzoom}
  }
\end{center}
\caption{Distribution of 1-minute returns and price changes from the Apollo Group Inc. (APOL) share in the first half of 2007. While Fig. (a) shows the distribution of the corresponding price changes, Fig. (b) shows the full distribution. Fig. (c) shows only the center of the distribution with the calculated bounds of corresponding to a specific price change indicated by the blue regions. Darker shades of blue imply overlapping bounds.}
\label{fig:returns}
\end{figure}

As the price change $\Delta S_{\Delta t}$ can only take values that are multiples of the tick-size $q$, its histogram consists of equally spaced peaks as shown in Fig. \ref{fig:dS}. In other words, the distribution of $\Delta S_{\Delta t}$ is discretized.
At first glance, it is conceivable that the transition from absolute price changes $\Delta S_{\Delta t}$ to relative price changes $r_{\Delta t}$ removes this discretization from the distribution, since the returns are almost continuously distributed, as Fig. \ref{fig:r} illustrates. However, a closer look at the center of the distribution in Fig. \ref{fig:rzoom} reveals that the discretization effects are still visible. Despite its non-visibility, the discretization affects returns on any interval. We will discuss this point more detailed in section \ref{s:corr}.

For an analytical description of this discretization, we introduce the set of all returns

\begin{equation}
R_{\Delta t}=\left\{\frac{ \Delta S_{\Delta t}(t) }{S(t)} \,\Big|\,\Delta S_{\Delta t}(t) \in \left[N_{-}q,(N_{-}+1)q, \dots, (N_{+}-1)q,N_{+}q\right]   \right\}\ ,
\end{equation}
where $N_{-}q$ defines the lower and $N_{+}q$ the upper bound of the price change distribution that is discretized by the tick-size $q$.

The set of all returns $R$ can be separated into subsets for each price change $\Delta S_{\Delta t}$, 
\begin{equation}
R_{\Delta t}= \bigcup_{n=N_{-}}^{N_{+}} R^{(n)}_{\Delta t}\ ,
\end{equation}
with
\begin{equation}
R^{(n)}_{\Delta t}=\left\{\frac{ \Delta S_{\Delta t}(t) }{S^{(n)}(t)}\, \Big|\, \Delta S_{\Delta t}(t) = nq  \right\}\ .
\end{equation}
$S^{(n)}$ in the denominator refers to the subset of starting prices that increase (or decrease) by $nq_{S}$ in the interval $\Delta t$.
Therefore, $R^{(n)}_{\Delta t}$ represents the returns that are based on the price change $nq$.
Evidently, $R^{(n)}_{\Delta t}$ is bounded by

\begin{equation}
\min(R^{(n)}_{\Delta t}) = \frac{nq}{\max(S^{(n)})}
\quad,\quad
\max(R^{(n)}_{\Delta t}) = \frac{nq}{\min(S^{(n)})} \ .
\label{eq:bounds}
\end{equation}
In our study, empirical data from the TAQ database \cite{TAQ} of the New York Stock exchange (NYSE) indicate that the approximations $\max(S^{(n)}) \approx \max(S)$ and $\min(S^{(n)}) 
\approx \min(S)$ are legitimate for small $|n|$.

Therefore, the interval between minimum and maximum return on a specific price change
\begin{equation}
I(R^{(n)}) = \left[\min\left(R^{(n)}\right),\max\left(R^{(n)}\right)\right]
\label{eq:rinterval}
\end{equation}
increases with $|n|$, while the distance $d$ between their centers remains almost constant
\begin{equation}
d(R^{(n)}) = \frac{q_{S}}{2}\left(\frac{1}{\min(S^{(n)})}-\frac{1}{\max(S^{(n)})}\right) \approx \frac{q_{S}}{2}\left(\frac{1}{\min(S)}-\frac{1}{\max(S)}\right) = \mathrm{const.}
\label{eq:rdistance}
\end{equation}
Thus, the intervals $I(R^{(n)})$ are increasingly overlapping for larger $|n|$. From this viewpoint the discretization is only ``visible'' for small $|n|$, that is, for small price changes. Fig. \ref{fig:rzoom} illustrates the clustering of returns with an example where we compare the returns of the Apollo Group Inc. (APOL) share with the intervals $I(R^{(n)})$ calculated by equations (\ref{eq:rinterval}) and (\ref{eq:rdistance}). The calculated boundaries match with the empirical data.


\subsection{Tail behavior of return and price change distribution}
\label{s:tails}
\begin{figure}[tb]
\begin{center}
\subfigure[gaussian $\Delta S$, $S_{\mathrm{max}}/S_{\mathrm{min}}=1.1$]{
  \includegraphics[width=0.31\textwidth]{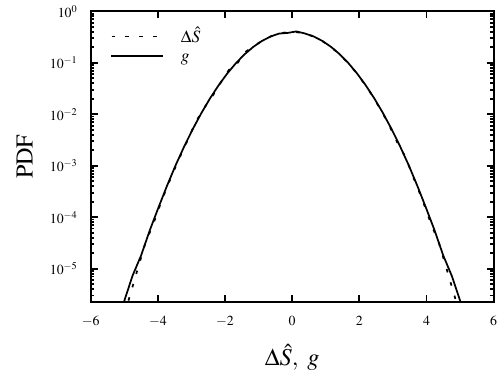}
  \label{fig:tailsg1}
  }
 \subfigure[gaussian $\Delta S$, $S_{\mathrm{max}}/S_{\mathrm{min}}=1.5$]{
   \includegraphics[width=0.31\textwidth]{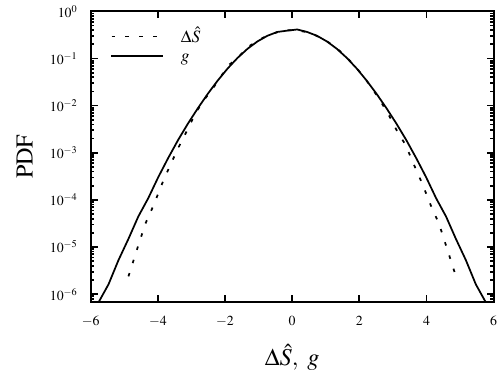}
     \label{fig:tailsg2}
  }
 \subfigure[gaussian $\Delta S$, $S_{\mathrm{max}}/S_{\mathrm{min}}=2.0$]{
   \includegraphics[width=0.31\textwidth]{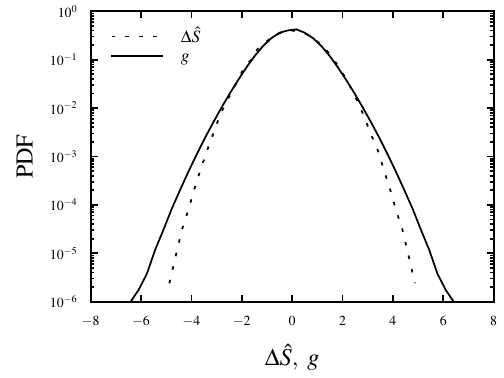}
     \label{fig:tailsg3}
  }\\
  \subfigure[powerlaw $\Delta S$, $S_{\mathrm{max}}/S_{\mathrm{min}}=1.1$]{
  \includegraphics[width=0.31\textwidth]{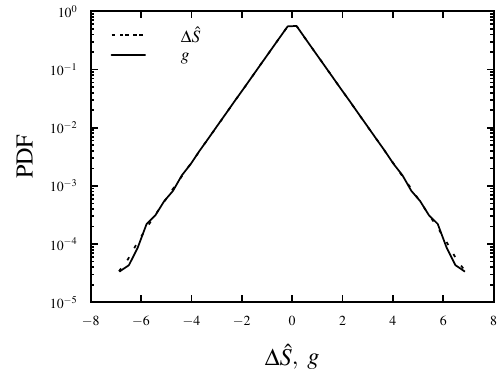}
  \label{fig:tailsp1}
  }
 \subfigure[powerlaw $\Delta S$, $S_{\mathrm{max}}/S_{\mathrm{min}}=1.5$]{
   \includegraphics[width=0.31\textwidth]{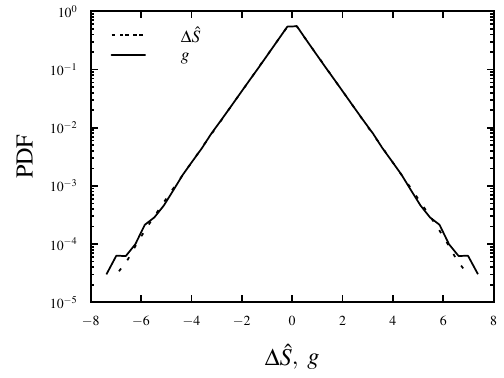}
     \label{fig:tailsp2}
  }
 \subfigure[powerlaw $\Delta S$, $S_{\mathrm{max}}/S_{\mathrm{min}}=2.0$]{
   \includegraphics[width=0.31\textwidth]{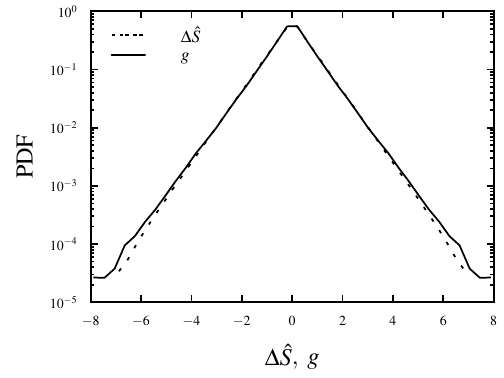}
     \label{fig:tailsp3}
  }
  
\caption{Comparision of the distributions of normalized price changes $\Delta \hat{S}$ and normalized returns $g$ on different price ranges $S_{\mathrm{max}}/S_{\mathrm{min}}$. Figs. (a), (b) and (c) have been calculated using Gaussian distributed price changes. Figs. (d), (e) and (f) have been calculated using exponential distributed prices. All calculations were preformed using a standard deviation of 60 tick-sizes.}
\label{fig:tails}
\end{center}
\end{figure}
We will now investigate, how the composition of the returns changes the shape of their distribution compared to the distribution of price changes. In the framework of a model, we generate price changes that are, in a first scenario, Gaussian distributed and, in a second scenario, powerlaw distributed with a given tick-size. Afterwards, we calculate returns using uniformly distributed price values within the regions $S_{\mathrm{min}}$ and $S_{\mathrm{max}}$ (analogously to Figs.\ref{fig:rzoom} and \ref{fig:r}).
In this manner, we generate a discrete price change distribution with a specific shape and then divide each set of equal price changes by uniformly distributed prices. The price distributions are generated individually for each subset.

To compare the shape of the obtained return distribution with the shape which we have chosen for the price change distribution, we normalize the distributions to zero mean and unit variance
\begin{equation}
g^{(i)}_{\Delta t}(t)=\frac{r^{(i)}_{\Delta t}(t)-\langle r^{(i)}_{\Delta t}  \rangle}{\sigma_{r^{(i)}_{\Delta t}}}
,\quad
\Delta\hat{S}^{(i)}_{\Delta t}(t)=\frac{\Delta S^{(i)}_{\Delta t}(t)-\langle \Delta S^{(i)}_{\Delta t}  \rangle}{\sigma_{\Delta S^{(i)}_{\Delta t}}}\ ,
\end{equation}
where $\langle \ldots \rangle$ denotes the mean value of a time series with length $T$ and where $\sigma$ refers to the standard deviation of the same time series. The index $i$ corresponds to the a certain security, e.g., a stock.

The results of this simple setup indicate that neither the tick-size nor the width of the price change distribution or the absolute sizes of $S_{\mathrm{min}}$ and $S_{\mathrm{max}}$ have an effect on the shape of the obtained return distribution. Only the microstructure of its center is affected, as discussed in the previous section.
In general, the return distribution acquires stronger tails compared to the price change distribution. Surprisingly, the shape-change of the distribution only depends on the ratio of the minimum and maximum price.

Figure \ref{fig:tails} shows the corresponding distributions for Gaussian and powerlaw distributed prices and for various price ranges. It turns out that the influence on the tail behavior is much stronger for a Gaussian price change distribution. For a powerlaw price change distribution, the return distribution retains almost the same powerlaw shape, except for the tails far out, while their center becomes slightly sharper.

\begin{figure}[tb]
\begin{center}
\subfigure[$\Delta t = 5\ \mathrm{min}$]
{
  \includegraphics[width=0.48\textwidth]{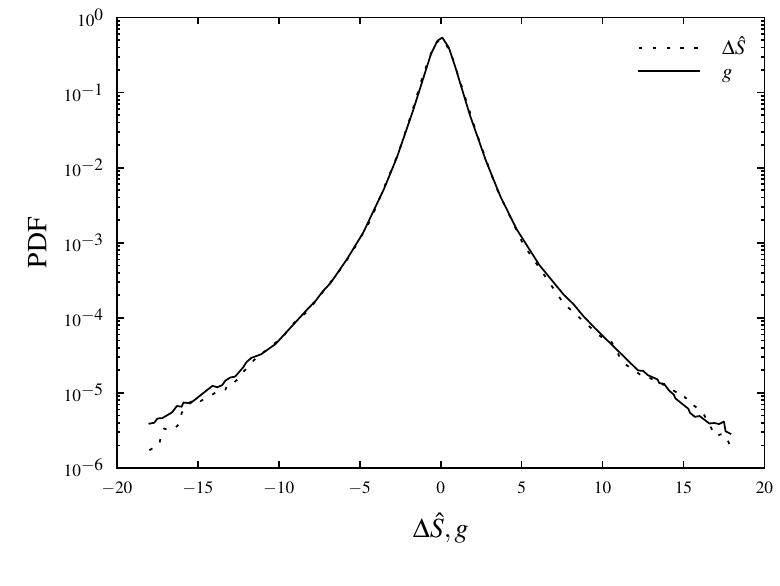}
 } 
\subfigure[$\Delta t = 1\ \mathrm{d}$]{
  \includegraphics[width=0.48\textwidth]{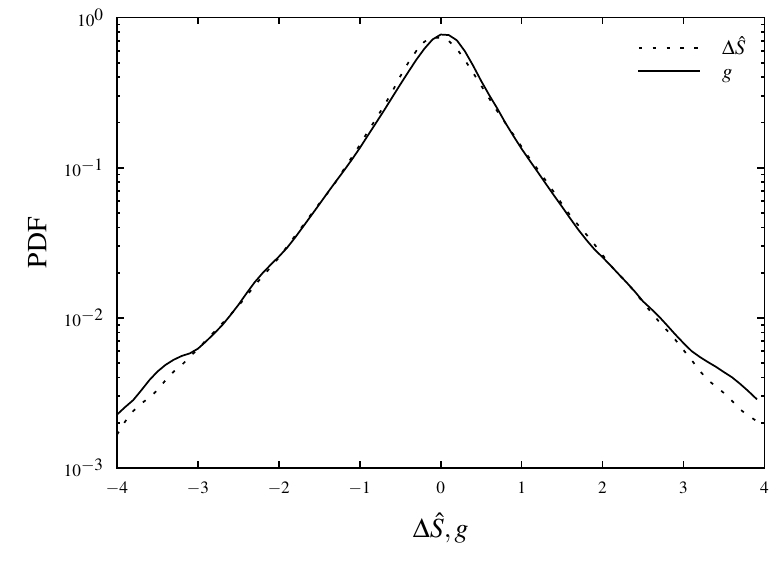}
 }
  
\caption{Change in the tail behavior of the distribution of normalized returns $g$ compared to the underlying normalized price changes $\Delta \hat{S}$. The price changes and returns have been calculated for an ensemble of 50 stocks from the S\&P 500 index using return intervals of 5 minutes and 1 day. The stocks were chosen to provide the highest relation between price variance and mean price.}
\label{fig:realtails}
\end{center}
\end{figure}

Of course, the assumption of uniformly distributed prices on each price change is a rough approximation within this simple setup. In the market, there can be a strong relation between $\Delta S$ and $S$, which leads to a shape retaining of the price change distribution to the return distribution. This is because the prices which undergo a very large price change during the interval $\Delta t$ can be much more sparsely distributed than prices which change only slightly. Furthermore, the price range is usually not very high in a period of time, in which the price distribution is approximately uniform. In view of this and under the assumption of powerlaw distributed price changes, the situation in Figs. \ref{fig:tailsp1} and \ref{fig:tailsp2} may describe most stocks suitably. Put differently, the shape of the return distribution is almost retaining the shape of the price change distribution in most cases.

However, if the price of a stock covers a large range in a relatively short period of time, we actually can observe a change in the tail behavior. This is illustrated in Fig. \ref{fig:realtails} for an ensemble of 50 stocks taken from the S\&P 500 index (See Tab. \ref{tab:var}). The stocks have been chosen to provide the highest ratio between their mean price and its standard deviation. Although the stock ensemble shows the expected behavior, it is difficult to make an accurate statement regarding the tails far out, as these events are very rare, even within this statistical ensemble.

\subsection{Tail behavior of the return microstructure}
\label{s:microtail}
\begin{figure}[tb]
\begin{center}
\includegraphics[width=0.48\textwidth]{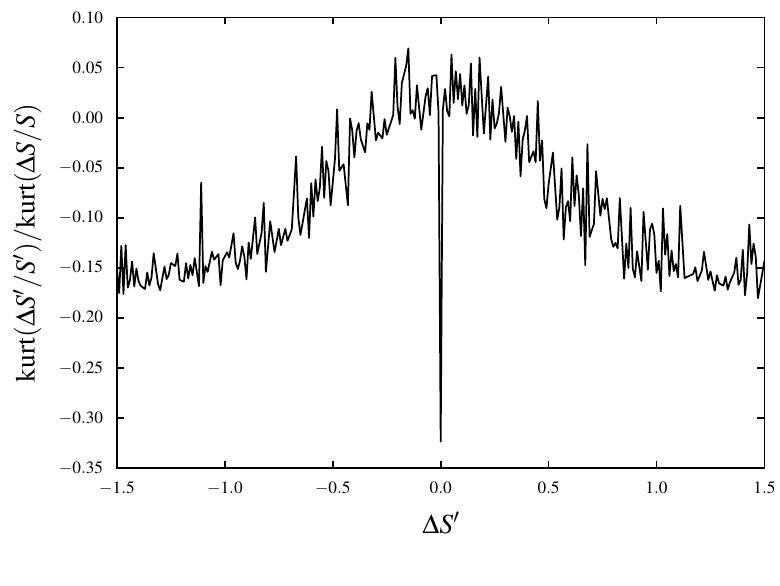}
\caption{Comparison of the kurtosis of the return distributions on a specific price change $\Delta S$ compared to the complete return distribution. The negative peak at $\Delta S' = 0$ originates from the fact that this return subset ``distribution'' of $\Delta S'/S$ only contains returns with the value zero and therefore leads to a value of $-3/\mathrm{kurt}(\Delta S/S)$.}
\label{fig:kurtosis}
\end{center}
\end{figure}

Another question arises in this regard: If the return distribution is heavy tailed, how is this connected to the tail behavior of the subsets of returns? As we demonstrated in the previous section, the set of all returns can be divided into subsets that are corresponding to a certain price change. Now, do these subparts feature stronger or weaker tails than the complete return distribution?

In Fig. \ref{fig:kurtosis}, we compare the kurtosis of the return subsets normalized to the kurtosis of the overall return distribution. As it is difficult to perform a proper normalization of a stock ensemble in this graphical representation, we show the result for the Google Inc. (GOOG) share as an example.

We make two observations: First, there seems to be a connection between the tail behavior and the return interval for the return subset distributions. The return subset distributions feature stronger tails for smaller price changes.
Second, surprisingly the return subset distributions exhibit a much smaller kurtosis than the complete return distribution. The strong tails of the complete distribution develop not until combining all the return subset distributions.

\section{Financial correlations}
\label{s:corr}
We now turn to the impact of the tick-size on the calculation of correlations and analyze the influence on the decay of correlation coefficients towards smaller return intervals (Epps effect).
Financial correlations are an important measure in economics. The knowledge of precise correlations is essential for quantifying and minimizing financial risk. As we will show, the discreteness of stock quotes can distort the calculated correlation coefficients.

A financial return is a compound observable value. Due to that fact, we develop the compensation method step by step.
We start in section \ref{a:quantcor}, where we turn to the distortion of the correlation coefficient of value-discretized time series in general. We develop a compensation for the discretization error in the correlation between financial (absolute) price changes in section \ref{s:error}. In section \ref{s:return_correct}, we extend this formalism to financial returns.

It is a basic assumption in our model that we can statistically describe the discreteness in market prices by a discretization of a hypothetical underlying continuous price. This is not to say that market prices actually result from a discretization process. Individual traders are well aware of the finite tick size and may try to exploit it in their trading strategies. However, there is a large variety of trading strategies simultaneously acting on the market.  
These strategies involve a large scale of different investment horizons. Since the price formation results from the interaction of a large diversity of strategies, the price fluctuations on the level of the tick size can be viewed as purely statistical. This is the basis for our modeling ansatz.

Despite the interpolation of the price change distribution, neither parameter fixing nor calibration of the model is necessary, in contrast to many other compensation methods for the Epps effect \cite{zebedee09,voev07,griffin06,corsi07,zhang08, barndorff08}.

\subsection{Calculation of the correlation coefficient for value-discretized time series}
\label{a:quantcor}
Almost any time series of data is discretized. This can simply be caused by numerical reasons, such as a finite number of decimal places. But how can we measure the impact of the discretization or even compensate it? We will show, that this can simply be achieved by a decomposition of the correlation coefficient and a estimation of the average discretization errors.

Let $x_{1}$ and $x_{1}$ be two time series which are correlated. The correlation coefficient of $x_{1}$ and $x_{2}$ is given by
\begin{equation}
\label{eq:basiccorr}
\mathrm{corr}(x_{1},x_{2})= \frac{\left\langle x_{1} x_{2} \right\rangle - \left\langle x_{1} \right\rangle \left\langle x_{2} \right\rangle}{\sigma_{1}\sigma_{2}}\ .
\end{equation}
Now we consider the time series $\bar{x}_{1}$ and $\bar{x}_{2}$ which are the discretized values of $x_{1}$ and $x_{2}$ with tick-sizes $q_{1}$ and $q_{2}$, respectively. Thus we have
\begin{eqnarray}
x_{1}(t)&=&\bar{x}_{1}(t)+\vartheta^{(1)}(t)\label{eq:discx} \\
x_{2}(t)&=&\bar{x}_{2}(t)+\vartheta^{(2)}(t)\label{eq:discy} \ ,
\end{eqnarray}
where $\vartheta^{(1)}(t)$ and $\vartheta^{(2)}(t)$ are the discretization errors.
We assume the discretization errors as uniformly distributed in the intervals $]-q_1/2,q_1/2]$ and $]-q_2/2,q_2/2]$. This seems natural, as a discretization is commonly caused by a rounding process.

Using equations (\ref{eq:discx}) and (\ref{eq:discy}) we can write the correlation coefficient (\ref{eq:basiccorr}) as
\begin{eqnarray}
\mathrm{corr}(x_{1},x_{2}) &=&
\frac{\left\langle (\bar{x}_{1}+\vartheta^{(1)})(\bar{x}_{2}+\vartheta^{(2)}) \right\rangle - \left(\left\langle \bar{x}_{1} \right\rangle+\left\langle \vartheta^{(1)} \right\rangle\right) \left(\left\langle \bar{x}_{2} \right\rangle+\left\langle \vartheta^{(2)} \right\rangle\right)}
{\sqrt{\mathrm{var}\left(\bar{x}_{1}+\vartheta^{(1)}\right)}\sqrt{\mathrm{var}\left(\bar{x}_{2}+\vartheta^{(2)}\right)}}\\
&=&\frac{\mathrm{cov}\left(\bar{x}_{1},\bar{x}_{2}\right)+\mathrm{cov}\left(\bar{x}_{1},\vartheta^{(2)}\right)+\mathrm{cov}\left(\bar{x}_{2},\vartheta^{(1)}\right)+\mathrm{cov}\left(\vartheta^{(1)},\vartheta^{(2)}\right)}
{\sqrt{\mathrm{var}\left(\bar{x}_{1}\right)+\mathrm{var}\left(\vartheta^{(1)}\right)+2\mathrm{cov}\left(\bar{x}_{1},\vartheta^{(1)}\right)}
\sqrt{\mathrm{var}\left(\bar{x}_{2}\right)+\mathrm{var}\left(\vartheta^{(2)}\right)+2\mathrm{cov}\left(\bar{x}_{2},\vartheta^{(2)}\right)}} \label{eq:errorcorr} \ .
\end{eqnarray}
Apart from the terms $\mathrm{cov}\left(\bar{x}_{1},\bar{x}_{2}\right)$, $\mathrm{var}\left(\bar{x}_{1}\right)$ and $\mathrm{var}\left(\bar{x}_{2}\right)$ of expression (\ref{eq:errorcorr}), which can be calculated with the discretized data, all other terms are lost in the discretization process. However, these terms can be estimated when the distributions $\varrho_{\bar{x}_{1}}$ and $\varrho_{\bar{x}_{2}}$ of $\bar{x}_{1}$ and $\bar{x}_{2}$ are known, as we will demonstrate. The continuous distributions $\varrho_{x_{1}}$ and $\varrho_{x_{2}}$ can be obtained by interpolating the distributions of the discretized values (we assume these distributions in the following context to be normalized).
Sometimes, the shape of the distribution for a certain process is known (e.g. Gaussian). Therefore, the interpolated distribution function can be determined by a fit of the distributions of $\bar{x}_{1}$ and $\bar{x}_{2}$. 

If the shape of the distribution is unknown, an interpolation can be performed section by section using e.g. polynomial or linear fits. The fitting processes cannot be performed as typically by minimizing the difference of values from the discrete distribution and the desired fit function. Rather the discretization process needs to be included. This gains particular importance when the level of discretization is high and thus the distribution is discretized only with a small range of values.

As the value that has been discretized to e.g. $x'_{1}$ can originate from region $x'_{1}-q_{1}/2$ to $x'_{1}+q_{1}/2$, the difference function $f$, which provides a measure for the residual between the fit and the empirical data is then given by

\begin{eqnarray}
f_{x_{1}}(\varrho_{x_{1}},\varrho_{\bar{x}_{1}})=\sum\limits_{n=N_-}^{N_+}
\left[\,
\int\limits_{q_1(n-\frac{1}{2})}^{q_1(n+\frac{1}{2})}\varrho_{x_{1}}(z)\,dz - 
\varrho_{\bar{x}_{1}}(nq_1)
\right]
\end{eqnarray}
for $x_{1}$ and analogously for $x_{2}$.

To compensate the overall discretization error, we first introduce the discretization errors that led to a certain discretized value. We call these errors conditional discretization errors. They are defined as

\begin{eqnarray}
\vartheta_{n}^{(1)}\left(\tilde{t}\right) &=& x_{1}(\tilde{t})-nq_{1}\ ,\ \tilde{t} \in \left\{t\:\: \big|\:\: |\:x_{1}(t)-nq_{1}|\le\frac{q_{1}}{2}\right\}\\
\vartheta_{m}^{(2)}\left(\tilde{t}\right) &=& x_{2}(\tilde{t})-mq_{2}\ ,\ \tilde{t} \in \left\{t\:\: \big|\:\: |x_{2}(t)-mq_{2}|\le\frac{q_{2}}{2}\right\}\\
\vartheta_{n,m}^{(1)}\left(\tilde{t}\right) &=& x_{1}(\tilde{t})-nq_{1}\ ,\ \tilde{t} \in \left\{t\:\: \big|\:\: |x_{1}(t)-nq_{1}|\le\frac{q_{1}}{2},|x_{2}(t)-mq_{2}|\le\frac{q_{2}}{2}\right\}\\
\vartheta_{m,n}^{(2)}\left(\tilde{t}\right) &=& x_{2}(\tilde{t})-mq_{2}\ ,\ \tilde{t} \in \left\{t\:\: \big|\:\: |x_{2}(t)-mq_{2}|\le\frac{q_{2}}{2},|x_{1}(t)-nq_{1}|\le\frac{q_{1}}{2}\right\}\ .
\end{eqnarray}
Here, $\vartheta_{n}^{(1)}$ and $\vartheta_{m}^{(2)}$ are the discretization errors that resulted in a discrete value of $\bar{x}_{1}=nq$ and $\bar{x}_{2}=mq$ accordingly, where $n$ and $m$ are integers. 
Consequently, $\vartheta_{n,m}^{(1)}$ and $\vartheta_{m,n}^{(2)}$ are discretization errors that led to a value of $\bar{x}_{1}=nq$ and $\bar{x}_{2}=mq$, while the other (correlated) time series was simultaneously discretized to $\bar{x}_{2}=mq$ and $\bar{x}_{1}=nq$. In all cases, $\tilde{t}$ is the set of time points at which these actual discretizations occur. 

Using the interpolated distribution functions $\varrho_{x_{1}}(x(t))$ and $\varrho_{x_{1}}(y(t))$ and the interpolated joint distribution function $\varrho_{x_{1},x_{2}}(x(t),y(t))$, the average discretization errors can be calculated as

\begin{eqnarray}
\left\langle \vartheta^{(1)}_{n} \right\rangle &=& \int_{q_{1}\left(n-\frac{1}{2}\right)}^{q_{1}\left(n+\frac{1}{2}\right)} (z-nq_{1})\varrho_{x_{1}}(z)\,dz
\,\Big/
\int_{q_{1}\left(n-\frac{1}{2}\right)}^{q_{1}\left(n+\frac{1}{2}\right)} \varrho_{x_{1}}(z)\,dz \\
\left\langle \vartheta^{(2)}_{m} \right\rangle &=& \int_{q_{2}\left(m-\frac{1}{2}\right)}^{q_{2}\left(m+\frac{1}{2}\right)} (z-mq_{2})\varrho_{x_{2}}(z)\,dz
\,\Big/
\int_{q_{2}\left(m-\frac{1}{2}\right)}^{q_{2}\left(m+\frac{1}{2}\right)} \varrho_{x_{2}}(z)\,dz \\
\left\langle \vartheta^{(1)}_{n,m} \right\rangle &=& \int_{q_{x}\left(n-\frac{1}{2}\right)}^{q_{1}\left(n+\frac{1}{2}\right)} (z-nq_{1})\varrho_{x_{1},y_{2}}(z,mq_{2})\,dz
\,\Big/
\int_{q_{1}\left(n-\frac{1}{2}\right)}^{q_{1}\left(n+\frac{1}{2}\right)} \varrho_{x_{1},x_{2}}(z,mq_{2})\,dz \\
\left\langle \vartheta^{(2)}_{m,n} \right\rangle &=& \int_{q_{2}\left(m-\frac{1}{2}\right)}^{q_{2}\left(m+\frac{1}{2}\right)} (z-mq_{2})\varrho_{x_{1},x_{2}}(nq_{1},z)\,dz
\,\Big/
\int_{q_{2}\left(m-\frac{1}{2}\right)}^{q_{2}\left(m+\frac{1}{2}\right)} \varrho_{x_{1},x_{2}}(nq_{1},z)\,dz \ ,
\end{eqnarray}
where 
\begin{eqnarray}
\int\limits_{-\infty}^{+\infty} \varrho_{x_{1},x_{2}}(x_{1}(t),z)\,dz &=& \varrho_{x_{1}}(x_{1}(t))\quad\mathrm{and}\\
\int\limits_{-\infty}^{+\infty} \varrho_{x_{1},x_{2}}(z,x_{2}(t))\,dz &=& \varrho_{x_{2}}(x_{2}(t))\ .
\end{eqnarray}
Therefore the overall average discretization errors can be written as
\begin{eqnarray}
\left\langle \vartheta^{(1)} \right\rangle &\approx& 
\frac{1}{T}\sum\limits_{n=N_{-}}^{N_{+}} T_{n} \left\langle \vartheta_{n}^{(1)}\right\rangle\\
\left\langle \vartheta^{(2)} \right\rangle &\approx& 
\frac{1}{T}\sum\limits_{m=M_{-}}^{M_{+}} T_{m} \left\langle \vartheta_{m}^{(2)}\right\rangle\ , \label{eq:univary}
\end{eqnarray}
where $T_{n}$ and $T_{m}$ are the number of values that have been discretized to $nq_1$ and $mq_2$.

Now we can calculate the discretization terms of equation (\ref{eq:errorcorr}). We begin with:
\begin{eqnarray}
\mathrm{cov}\left(\bar{x}_{1},\vartheta^{(2)}\right) &=& 
\left\langle \bar{x}_{1} \vartheta^{(2)} \right\rangle - \left\langle \bar{x}_{1} \right\rangle \left\langle \vartheta^{(2)} \right\rangle \\
&=&
\frac{1}{T} \sum\limits_{n=N_{-}}^{N_{+}}\sum\limits_{m=M_{-}}^{M_{+}}\sum\limits_{\tilde{t}=0}^{T_{n,m}}\left(nq_{1} \vartheta_{m}^{(2)}(\tilde{t})\right)
- \left\langle \bar{x}_{1} \right\rangle
\left\langle \vartheta^{(2)} \right\rangle\\
&=&
\frac{q_{1}}{T} \sum\limits_{n=N_{-}}^{N_{+}} n \sum\limits_{m=M_{-}}^{M_{+}} T_{n,m}
\left\langle \vartheta_{m,n}^{(2)}\right\rangle
- \left\langle \bar{x}_{1} \right\rangle
\left\langle \vartheta^{(2)} \right\rangle\
\label{eq:cov_x_theta_y} \ .
\end{eqnarray}
Here, $q_{1}N_{-}$ represents the minimum of the discretized time series $\bar{x}_{1}(t)$. $q_{1}N_{+}$ is its maximum. $T$ is the length of the whole time-series, while $T_{n,m}$ is the number of synchronous pairs of both time series, which are discretized to $nq_{1}$ and $mq_{2}$. We index these pairs with $\tilde{t}$ referring to these certain point in time.

Analogously, the other discretization terms of equation (\ref{eq:errorcorr}) can be calculated as 
\begin{eqnarray}
\mathrm{cov}\left(\bar{x}_{2},\vartheta^{(1)}\right)
&= &
\frac{q_{2}}{T} \sum\limits_{m=M_{-}}^{M_{+}} m \sum\limits_{n=N_{-}}^{N_{+}} T_{n,m}
\left\langle \vartheta_{n,m}^{(1)}\right\rangle \label{eq:cov_y_theta_x}
- \left\langle \bar{x}_{2} \right\rangle
\frac{1}{T}\sum\limits_{n=N_{-}}^{N_{+}} T_{n} \left\langle \vartheta_{n}^{(x)}\right\rangle
\\
\mathrm{cov}\left(\bar{x}_{1},\vartheta^{(1)}\right) &=&
\frac{q_1}{T}\sum\limits_{n=N_{-}}^{N_{+}} T_{n} n \left\langle \vartheta_{n}^{(1)}\right\rangle
- \left\langle \bar{x}_{1} \right\rangle
\frac{1}{T}\sum\limits_{n=N_{-}}^{N_{+}} T_{n} \left\langle \vartheta_{n}^{(1)}\right\rangle
\\
\mathrm{cov}\left(\bar{x}_{2},\vartheta^{(2)}\right) &=&
\frac{q_2}{T}\sum\limits_{m=M_{-}}^{M_{+}} T_{m} m \left\langle \vartheta_{m}^{(2)}\right\rangle
- \left\langle \bar{x}_{2} \right\rangle
\frac{1}{T}\sum\limits_{m=M_{-}}^{M_{+}} T_{m} \left\langle \vartheta_{m}^{(2)}\right\rangle
\\
\mathrm{var}\left(\vartheta^{(1)}\right) &\approx& \frac{q_1^2}{12} \label{eq:var_x}
\\
\mathrm{var}\left(\vartheta^{(2)}\right) &\approx& \frac{q_2^2}{12} \label{eq:var_y}
\end{eqnarray}
The terms (\ref{eq:var_x}) and (\ref{eq:var_y}) are estimated under the assumption that the discretization errors are uniformly distributed.
Usually, the remaining term $\mathrm{cov}(\vartheta^{(1)}_{n},\vartheta^{(2)}_{m})$ cannot be calculated with the distribution functions as it contains the correlation between the discretization errors. This value is not necessarily connected to the correlation of the whole time series either. Yet, we will show in the next section, that this term is negligible in the present context. 

Thus, we have shown that the error caused by the discretization can be estimated by decomposing the correlation coefficient and approximating the mean discretization errors by interpolating the discrete distributions.

\subsection{Distortion of price change correlations}
\label{s:error}
We now turn to the specific situation on the stock market. The situation differs, when applying the method from the previous section to stock price changes. Here, the discretization process does not take place on the actual observable. Instead the price change $\Delta S$ is a difference for two prices $S(t)$ and $S(t+\Delta t)$ that are discretized by the tick-size $q$.

\begin{figure}
\centering
	\subfigure[]
{
  \includegraphics[width=0.48\textwidth]{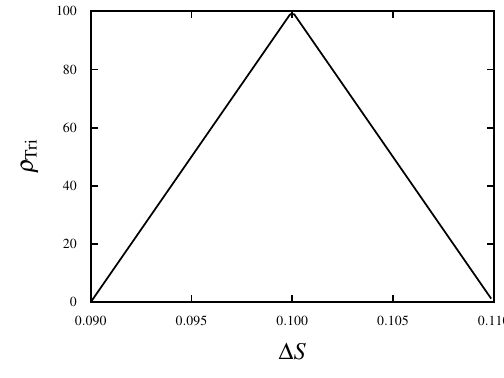}
  \label{fig:tri}
 } 	
 \subfigure[]
{
  \includegraphics[width=0.48\textwidth]{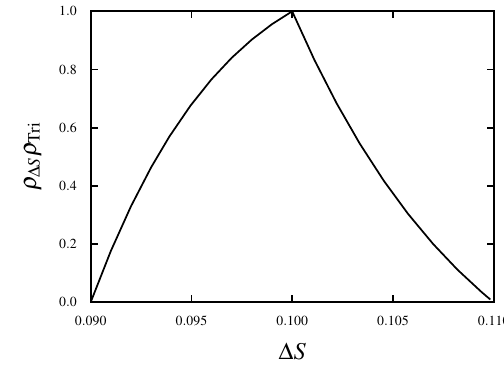}
  \label{fig:triplaw}
 } 
	\caption{Exemplary distribution of discretization errors around a price change of $\Delta S = 0.1$ and a tick-size of $q_{\Delta S}=0.01$ (a). Fig. (b) shows the product with power law distribution given by $\rho_{\Delta S}(x)=10x^{-6}$}

\end{figure}

Therefore, the discretization error on a specific price difference $\Delta S'$ can be in the range from $-q$ to $q$. However, the probability that a certain value is from a price difference within this range is not constant. It is described by a triangular-shaped distribution (See Fig. \ref{fig:tri}). This is evident, as the distribution error is the difference of two uniformly distributed discretization errors. The normalized triangular distribution $\varrho_{\mathrm{Tri}}$ around a certain price change $\Delta S'$ vanishes at $\Delta S'-q$ and $\Delta S'+q$ and has the value $1/q$ at its maximum at $\Delta S'$. It reads as

\begin{eqnarray}
\varrho_{\mathrm{Tri}}(x,\Delta S') = 
\begin{cases}
\frac{x-\Delta S'+q}{q^2} & (\Delta S' -q) \leq x < \Delta S'\\
\frac{-x+\Delta S'+q}{q^2} & (\Delta S' +q) \geq x \geq \Delta S'\\
0 & \mathrm{else}\ .
\end{cases}
\label{eq:tri}
\end{eqnarray}
The average discretization errors have now to be calculated with the product of the triangular distribution $\varrho_{\mathrm{Tri}}$ and the interpolated price change distributions $\varrho_{\Delta S_{1}}$, $\varrho_{\Delta S_{2}}$ (and proper normalization). Thus,

\begin{eqnarray}
\left\langle \vartheta^{(1)}_{n} \right\rangle &=& \int_{q_1\left(n-1\right)}^{q_1\left(n+1\right)} (z-nq_{1})\varrho_{\Delta S_{1}}(z)\varrho_{\mathrm{Tri}}(z,nq_{1})\,dz
\,\Big/
\int_{q_{1}\left(n-1\right)}^{q_{1}\left(n+1\right)} \varrho_{\Delta S_{1}}(z)\varrho_{\mathrm{Tri}}(z,nq_{1})\,dz
\label{eq:meantriplaw}\ ,\\
\left\langle \vartheta^{(1)}_{n,m} \right\rangle &=& \int_{q_1\left(n-1\right)}^{q_1\left(n+1\right)} (z-nq_{1})\varrho_{\Delta S_{1},\Delta S_{2}}(z,mq_{2})\varrho_{\mathrm{Tri}}(z,nq_{1})\,dz
\,\Big/
\int_{q_1\left(n-1\right)}^{q_1\left(n+1\right)} \varrho_{\Delta S_{1},\Delta S_{2}}(z,mq_{2})\varrho_{\mathrm{Tri}}(z,nq_{1})\,dz\ .\label{eq:meantriplawv}\
\end{eqnarray}
$\left\langle \vartheta^{(2)}_{m} \right\rangle$ and $\left\langle \vartheta^{(2)}_{m,n} \right\rangle$ are analogously defined.

Fig. \ref{fig:triplaw} shows exemplarily the product of a triangular distribution and a power law distribution. The denominator in equation (\ref{eq:meantriplaw}) refers to the area under this curve.
The triangular distribution also needs to be included in the fitting process. Thus, the difference function becomes

\begin{eqnarray}
f_{\Delta S}(\varrho_{\Delta S},\varrho_{\Delta \bar{S}})&=&\sum\limits_{n=N_-}^{N_+}
\left[\,
\int\limits_{q_{S}(n-1)}^{q_{S}(n+1)}\varrho_{\mathrm{Tri}}(z,nq_{S})\left[\varrho_{\Delta S}(z) - 
\varrho_{\Delta \bar{S}}(nq_{S})\right]\,dz
\right]\\
&=&\sum\limits_{n=N_-}^{N_+}
\left[\,
\int\limits_{q_{S}(n-1)}^{q_{S}(n+1)}\varrho_{\mathrm{Tri}}(z,nq_{S})\varrho_{\Delta S}(z)\,dz - 
\varrho_{\Delta \bar{S}}(nq_{S})
\right]\ .
\end{eqnarray}
Where $\varrho_{\Delta \bar{S}}$ refers to the discretized distribution. $\varrho_{\mathrm{Tri}}$ acts like a weighting function in the residual measure. It provides a weight corresponding to the probability that the difference of the originating discretization errors result in the value $z$.

Now, we are able to estimate the correlation discretization error with the previously defined equations (\ref{eq:cov_x_theta_y}) to (\ref{eq:var_y}).

\subsection{Distortion of return correlations}
\label{s:return_correct}
When calculating the correlation of financial returns as defined in equation (\ref{eq:returndef}) the situation becomes more complex. Here, we also have to take the prices into account. The correlation coefficient (\ref{eq:errorcorr}) for two return time series $r_1$ and $r_2$ now reads as 
\begin{equation}
\mathrm{corr}(r_1,r_2)
\frac{\mathrm{cov}\left(\bar{r}_1,\bar{r}_2\right)+
\mathrm{cov}\left(\frac{\Delta \bar{S}_1}{S_1},\frac{\vartheta^{(2)}}{S_2}\right)+\mathrm{cov}\left(\frac{\Delta \bar{S}_2}{S_2},\frac{\vartheta^{(1)}}{S_1}\right)+
\mathrm{cov}\left(\frac{\vartheta^{(1)}}{S_1},\frac{\vartheta^{(2)}}{S_2}\right)}
{\sqrt{\mathrm{var}\left(\bar{r}_1\right)+\mathrm{var}\left(\frac{\vartheta^{(1)}}{S_1}\right)+2\mathrm{cov}\left(\frac{\Delta \bar{S}_1}{S_1},\frac{\vartheta^{(1)}}{S_1}\right)}
\sqrt{\mathrm{var}\left(\bar{r}_2\right)+\mathrm{var}\left(\frac{\vartheta^{(2)}}{S_1}\right)+2\mathrm{cov}\left(\frac{\Delta \bar{S}_1}{S_1},\frac{\vartheta^{(1)}}{S_1}\right)}}\ .\label{eq:errorcorr_2}
\end{equation}
Here, $\bar{r}_1$ and $\bar{r}_2$ refer to the discretized return time-series.
Analogously to the correlation between price changes, the individual terms can be estimated, but in addition, the starting prices $S_{1}$ and $S_{2}$ need to be parameterized. We use the variables $k$ and $l$ for this. $q_{1}K_{-}$ represents the minimum price within the observed time series, while $q_{1}K_{+}$ represents the maximum price. $T_{n,m,k,l}$ represents the number of pairs whose returns equal $(q_{1}n)/(q_{1}k) =n/k$ and $m/l$. Similar to that, $T_{n,k}$ refers to the number of returns (from a single time-series) that are equal to $n/k$. Thus, we obtain
\begin{eqnarray}
\mathrm{cov}\left(\frac{\Delta \bar{S}_1}{S_1},\frac{\vartheta^{(2)}}{S_2}\right)
&\approx& \frac{q_{1}}{T}
\sum\limits_{n=N_{-}}^{N_{+}} n
\sum\limits_{m=M_{-}}^{M_{+}} q_{2}
\sum\limits_{k=K_{-}}^{K_{+}}
\sum\limits_{l=L_{-}}^{L_{+}}
T_{n,m,k,l}
\frac{\left\langle \vartheta_{m,n}^{(2)}\right\rangle}{kl}
- \left\langle\frac{\Delta \bar{S}_1}{S_1}\right\rangle
\left\langle\frac{ \vartheta^{(2)} }{S_{2}}\right\rangle
\label{eq:compstart}
\\
\mathrm{cov}\left(\frac{\Delta \bar{S}_2}{S_2},\frac{\vartheta^{(1)}}{S_1}\right)
&\approx& \frac{q_{2}}{T}
\sum\limits_{m=M_{-}}^{M_{+}} m
\sum\limits_{n=N_{-}}^{N_{+}} q_{1}
\sum\limits_{k=K_{-}}^{K_{+}}
\sum\limits_{l=L_{-}}^{L_{+}}
T_{n,m,k,l}
\frac{\left\langle \vartheta_{n,m}^{(1)}\right\rangle}{kl}
- \left\langle\frac{\Delta \bar{S}_2}{S_2}\right\rangle
\left\langle\frac{ \vartheta^{(1)} }{S_{1}}\right\rangle
\\
\mathrm{cov}\left(\frac{\Delta \bar{S}_1}{S_1},\frac{\vartheta^{(1)}}{S_1}\right) &\approx& 
\frac{q_{1}}{T}\sum\limits_{n=N_{-}}^{N_{+}} \sum\limits_{k=K_{-}}^{K_{+}} T_{n,k} \frac{n}{k^2} \left\langle \vartheta_{n}^{(1)}\right\rangle
- \left\langle\frac{\Delta \bar{S}_1}{S_1}\right\rangle
\left\langle\frac{ \vartheta^{(1)} }{S_{1}}\right\rangle
\label{eq:compmid}
\\
\mathrm{cov}\left(\frac{\Delta \bar{S}_2}{S_2},\frac{\vartheta^{(2)}}{S_2}\right) &\approx&
\frac{q_{2}}{T}\sum\limits_{n=N_{-}}^{N_{+}} \sum\limits_{k=K_{-}}^{K_{+}} T_{n,k} \frac{n}{k^2} \left\langle \vartheta_{n}^{(2)}\right\rangle
- \left\langle\frac{\Delta \bar{S}_2}{S_2}\right\rangle
\left\langle\frac{ \vartheta^{(2)} }{S_{2}}\right\rangle
\label{eq:compmid2}
\\
\mathrm{var}\left(\frac{\vartheta^{(1)}}{S_{1}}\right) &\approx& \frac{q_{1}^2}{6}
\left\langle \frac{1}{S_1^2}\right\rangle
\\
\mathrm{var}\left(\frac{\vartheta^{(2)}}{S_{2}}\right) &\approx& \frac{q_{2}^2}{6}
\left\langle \frac{1}{S_2^2}\right\rangle\ .\label{eq:compend}
\end{eqnarray}
The terms $\left\langle \vartheta^{(1)}/ S_{1}\right\rangle$ and analogously $\left\langle \vartheta^{(2)}/ S_{2}\right\rangle$ in equations (\ref{eq:compstart}) to (\ref{eq:compmid2}) can be estimated as
\begin{eqnarray}
\left\langle\frac{ \vartheta^{(1)} }{S_{1}}\right\rangle \approx
\frac{1}{T}
\sum\limits_{n=N_{-}}^{N_{+}} q_{1}
\sum\limits_{k=K_{-}}^{K_{+}}
T_{n,k}
\frac{\left\langle \vartheta_{n}^{(1)}\right\rangle}{k}\ .
\end{eqnarray}
We note that the correlation between $\Delta S$ and $S$ is neglected in this approximation. Also the discretization of the prices in the denominator of the return is not compensated. However, the model results in the next section demonstrate that this simplification only induces a minor error.

Also the impact of specific trading strategies can be calculated using the presented modeling. Here, the distortion of correlation coefficients, the distribution of discretization errors (equation (\ref{eq:tri})) needs to be chosen in a suitable manner.

\begin{figure}[p]
\begin{center}
\subfigure[$c=0.2$, $S^{(1)}_{t=0}=1000$, $S^{(2)}_{t=0}=1000$]{
  \includegraphics[width=0.48\textwidth]{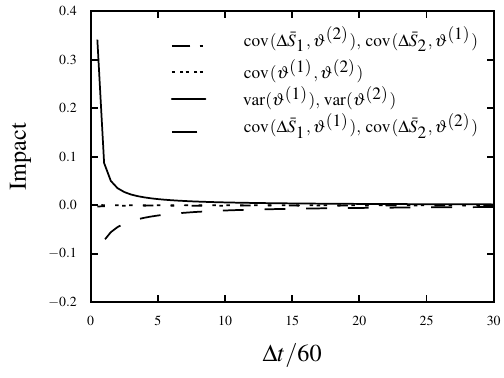}
  }
\subfigure[$c=0.4$, $S^{(1)}_{t=0}=1000$, $S^{(2)}_{t=0}=1000$]{
   \includegraphics[width=0.48\textwidth]{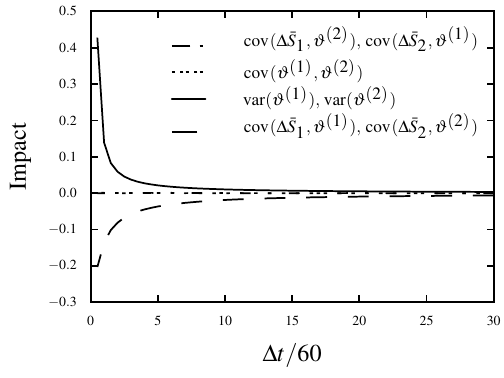}
  }\\
\subfigure[$c=0.8$, $S^{(1)}_{t=0}=1000$, $S^{(2)}_{t=0}=1000$]{
   \includegraphics[width=0.48\textwidth]{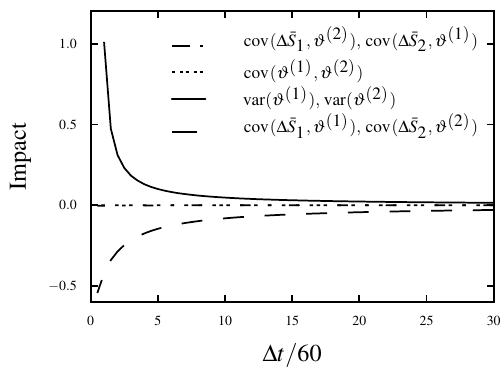}
  }
 \subfigure[$c=0.2$, $S^{(1)}_{t=0}=1000$, $S^{(2)}_{t=0}=10000$]{
  \includegraphics[width=0.48\textwidth]{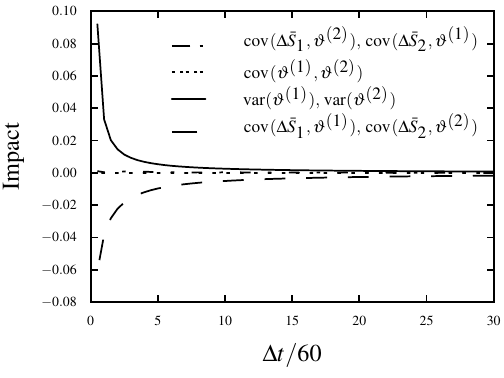}
  }\\
\subfigure[$c=0.4$, $S^{(1)}_{t=0}=1000$, $S^{(2)}_{t=0}=10000$]{
   \includegraphics[width=0.48\textwidth]{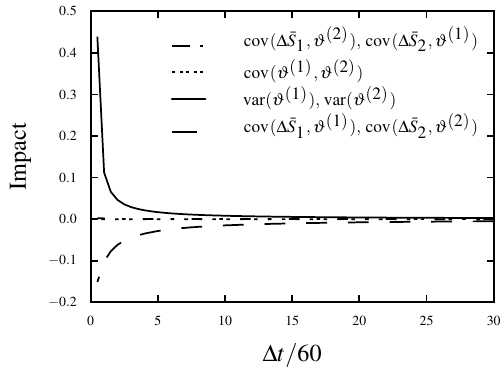}
  }
\subfigure[$c=0.8$, $S^{(1)}_{t=0}=1000$, $S^{(2)}_{t=0}=10000$]{
   \includegraphics[width=0.48\textwidth]{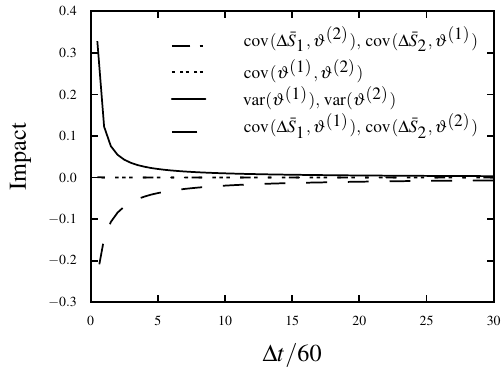}
  }
\caption{Impact of each term of the compensation method for the correlation coefficient between price changes.}
\label{fig:termsize}
\end{center}
\end{figure}

\begin{figure}
\centering
  \includegraphics[width=0.48\textwidth]{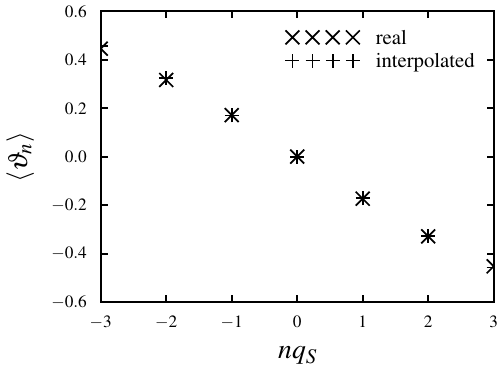}
  \caption{Benchmark of the error estimation: Comparison between real and estimated discretization errors within the model setup.}
  \label{fig:meanbench}
\end{figure}

\subsection{Results / Impact on the Epps effect}
\label{s:epps}
After we developed a method for compensating the discretization error in the calculation of correlations, we verify it in a model setup and apply it to empirical data.
We perform the presented compensation for different time intervals, in order to examine wether there is also a connection to the Epps effect. 

The Epps effect refers to the decay of the correlation coefficient towards small return intervals. 
Therefore financial correlations on returns which are based on intervals below a certain limit (e.g. 30 minutes) are unreliable. The ability to calculate the correlation structure on small return intervals is equivalent to an improved statistical significance or the gain of more recent information.
In previous studies the asynchrony of the time series has been identified as a major cause for the Epps effect \cite{muennix09b, hayashi05}. The following demonstrates that the price discretization can result in a sizable contribution to the Epps effect as well.

As the mean price change per return interval decreases with the length of the interval \cite{bouchaud04}, the width of the price change distribution decreases as well. While the tick-size remains constant, the discretization error increases. Hence, the tick-size should also have an impact on the Epps effect \-- especially for stocks which are traded at low prices.

\subsubsection{Model results}

Before applying the method to estimate the discretization error in empirical data, we evaluate it in a model setup. In addition, we will use the model to analyze the impact of each term from the decomposed correlation coefficient on the compensation.

We begin with generating an underlying correlated time series using the \emph{Capital Asset Pricing Model} (CAPM) \cite{sharpe64}, which is in a one-factor form known as Noh's model \cite{noh00} in physics,
\begin{equation}
r^{(i)}(t) = \sqrt{c}\,\eta(t) + \sqrt{1-c}\,\varepsilon^{(i)}(t)\ .
\end{equation}
Here $r^{(i)}$ stands for the return of the $i$-th stock at time $t$ and $c$ is the correlation coefficient. The random variables $\eta$ and $\varepsilon ^{(i)}$ are taken from standard normal distributions. Two return time series $r^{(1)}$ and $r^{(2)}$ are generated representing two correlated stocks. The lengths of these  time series is chosen as $7.2\cdot10^{6}$, corresponding to a return interval $\Delta t$ of 1 second during 1 trading year.

\begin{figure}[p]
\begin{minipage}[t]{0.48\textwidth}
\subfigure[$S^{(1)}_{t=0}=1000$, $S^{(2)}_{t=0}=1000$, $c=0.4$]{
  \includegraphics[width=1\textwidth]{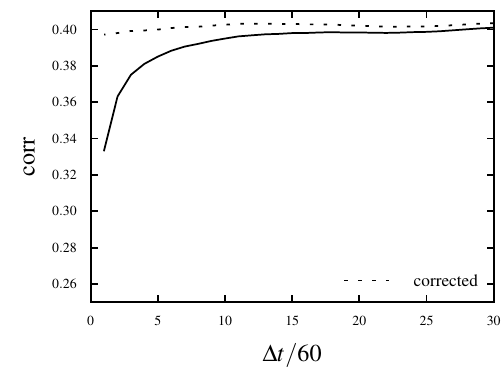}
  \label{fig:e11}
  }\\
\subfigure[$S^{(1)}_{t=0}=1000$, $S^{(2)}_{t=0}=10000$, $c=0.4$]{
   \includegraphics[width=1\textwidth]{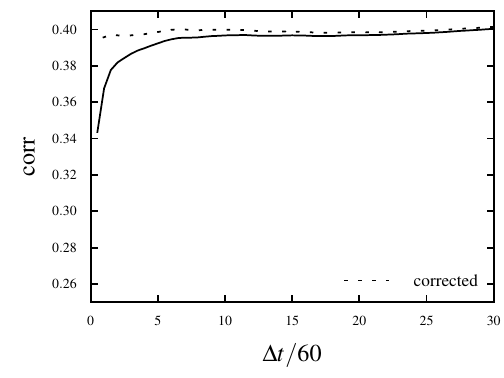}
     \label{fig:e110}
  }\\
\subfigure[$S^{(1)}_{t=0}=10000$, $S^{(2)}_{t=0}=10000$, $c=0.4$]{
   \includegraphics[width=1\textwidth]{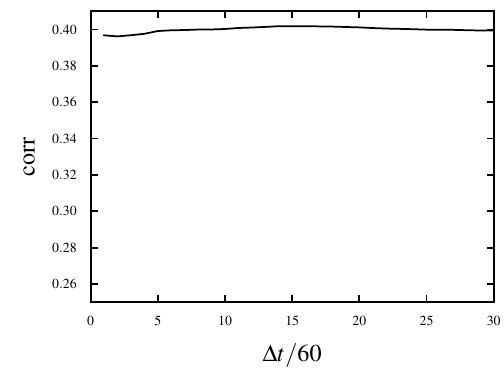}
     \label{fig:e1010}
  }
\caption{Left: Scaling behavior of the correlation coefficient of price changes due to the discretization error in the model setup. The dashed line represents the presented correction.}
\label{fig:res_ds}
\end{minipage}
\begin{minipage}[t]{0.04\textwidth}
\hfill
\end{minipage}
\begin{minipage}[t]{0.48\textwidth}
\subfigure[$S^{(1)}_{t=0}=1000$, $S^{(2)}_{t=0}=1000$, $c=0.4$]{
  \includegraphics[width=1\textwidth]{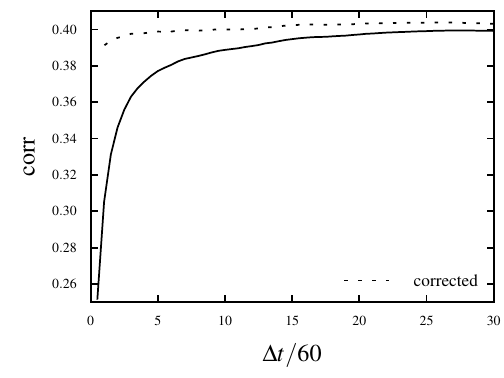}
  }\\
\subfigure[$S^{(1)}_{t=0}=1000$, $S^{(2)}_{t=0}=10000$, $c=0.4$]{
   \includegraphics[width=1\textwidth]{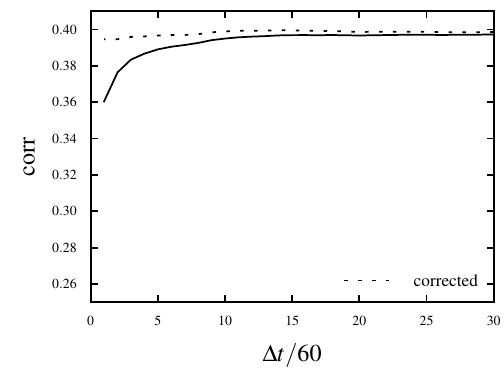}
  }\\
\subfigure[$S^{(1)}_{t=0}=10000$, $S^{(2)}_{t=0}=10000$, $c=0.4$]{
   \includegraphics[width=1\textwidth]{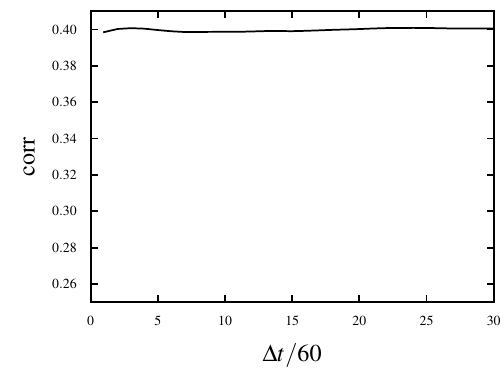}
  }
\caption{Right: Scaling behavior of the correlation coefficient of returns due to the discretization error in the model setup. The dashed line represents the presented correction.}
\label{fig:res_r}
\end{minipage}
\end{figure}

Using these returns, we generate two price time series $S^{(1)}$ and $S^{(2)}$ that perform a geometric Brownian motion with zero drift and a standard deviation of $10^{-3}$ per time step. 
The initial starting prices $S_{t=0}$ were set to $1000$ and $10000$.
In the next step, we round the prices to integer values. An integer price of for example 1000 then corresponds to a price of 10 and a tick-size of 0.01.

Now we are able to construct the discretized return time series $\bar{r}^{(i)}$ from these discretized prices using return intervals from 60 data points (corresponding to 1 minute) to 1800 data points (corresponding to 30 minutes).

As we know the actual discretization errors in the model, we can use it to evaluate our error estimates. A comparison of the estimated average discretization errors with the actual discretization errors is shown in Fig. \ref{fig:meanbench}. The estimated values show an excellent agreement with the original values. We restrict the interpolation to a single Gaussian fit, as we know the type of the price change distributions in this case. Thus, we can verify the scope of the estimation itself, not the suitability of the interpolation.

Before we perform the compensation, we want to see how much impact each correction term (equations (\ref{eq:compstart}) to (\ref{eq:compend})) has. We quantify the impact by calculating equation (\ref{eq:errorcorr_2}) and subtract the value of this expression with the regarded term set to zero. By this method we can see how the correlation coefficient changes, if a certain term of the discretization compensation is neglected (set to zero). Figure \ref{fig:termsize} illustrates the results of this analysis for different start prices and correlation coefficients. It turns out that only equations (\ref{eq:compmid}) to (\ref{eq:compend}) provide a sizable contribution to the compensation. Therefore, we restrict our compensation to the calculation of these terms. This implies that the distortion of the correlation coefficient is mainly caused by an improper normalization of the returns, as the terms (\ref{eq:compmid}) to (\ref{eq:compend}) only appear in the correction of the standard deviations of each return.

Thereby, we are able to compensate the discretization effects. We first focus on the correlations between price changes. As shown in Fig. \ref{fig:e11} and \ref{fig:e110}, the correlation coefficient decays towards smaller price change intervals.  Therefore, this effect is also a cause of the Epps effect. This effect becomes especially relevant when the ratio of the price to the tick-size is sufficiently small. It is remarkable that this scaling behavior is observed even though the time series are synchronous. The effect vanishes in our simulation, when both prices start with a value of 10000, as Fig. \ref{fig:e1010} illustrates. 

When applying the compensation method to return time series as illustrated in Fig. \ref{fig:res_r}, we are also able to correct the discretization error almost completely. The slight decay of the corrected correlation coefficient on very small return intervals is due to approximations, as stated at the end of section \ref{s:return_correct}. These are the negligence of the correlation between price changes and prices. In addition, even though the discretization of price changes is corrected, the price discretization in the denominator of the return is neglected. A further improvement of the compensation could be achieved by including these effects. However this would require further assumptions on the price process and would increase the necessary computing time dramatically. Thus, we restrict ourselves to the presented compensation.

\subsubsection{Empirical results}
\begin{figure}[tb]
\begin{center}
\subfigure[Ensemble: \$0.01--\$10.00]{
  \includegraphics[width=0.48\textwidth]{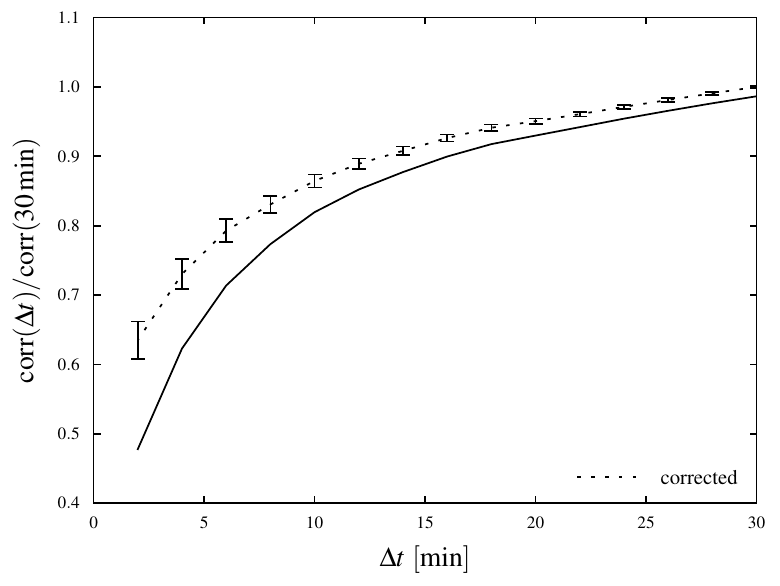}
  \label{fig:epps1}
  }
 \subfigure[Ensemble: \$10.01--\$20.00]{
   \includegraphics[width=0.48\textwidth]{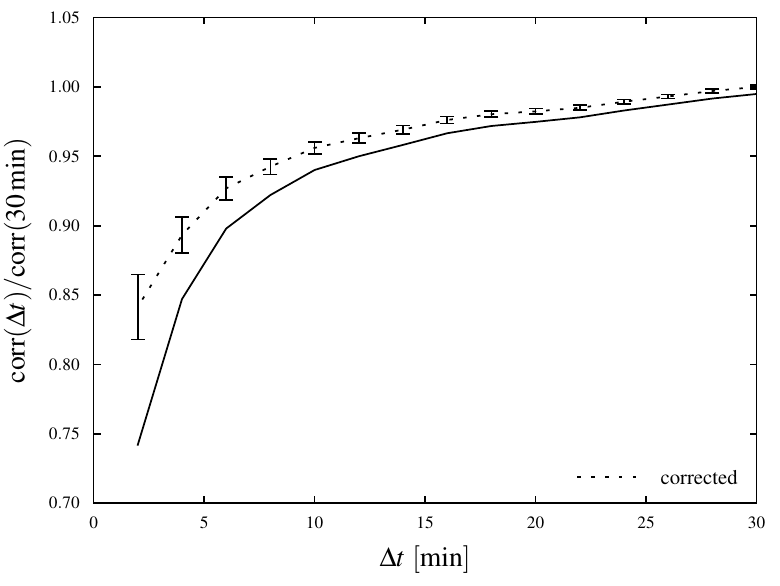}
     \label{fig:epps2}
  }
  
\caption{Tick-size compensation of the correlation coefficient between two ensembles consisting of the 25 highest correlated stocks from the S\&P 500 index that are averagely quoted within the region of \$0.01-\$10.00 and \$10.01-\$20.00, respectively. The correlation coefficients have been normalized to its saturation value at approximately 30 min. The plot the error bars represents the double standard deviation $2\sigma$. The  correlation coefficients for a return interval of 30 minutes averages to 0.19 (a) and 0.37 (b).}
\label{fig:realepps}
\end{center}
\end{figure}
How large is the contribution of the discretization effect to the Epps effect? To answer this, we apply the compensation to empirical data from the NYSE TAQ database \cite{TAQ}. 
Here, we use a powerlaw approach for the interpolation of the price change distribution, as the model results indicate that the discretization effects are mainly relevant for small return intervals. On small return intervals, powerlaw tails can describe the distribution satisfactory \cite{gopikrishnan99}. We perform a least squares fit of $a$ and $b$ in $\varrho_{\Delta S} = ax^{-|b|}$ for each value of the (discrete) distribution and their next two left and right neighbors individually. For the very central part of the distribution, a Gaussian fit was performed.

It is particularly important that stock splits must not be corrected in order to maintain the correct tick-size. Of course, therefore overnight returns have to be excluded. To analyze the impact of the discretization effect, we construct two ensembles (See Tab. \ref{tab:high1} and \ref{tab:high2}) of stocks from the S\&P 500 index. The first ensemble consists of stocks that are averagely priced between \$0.01 and \$10.00. The second ensemble consists of stocks that are on average priced between \$10.01 and \$20.00. Both ensembles are composed of 25 stock pairs providing the highest correlation during the year 2007 (based on daily data). 

As figure \ref{fig:realepps} demonstrates, we are able to compensate the impact of the tick-size on the correlation coefficient in empirical market data. Certainly, the decay can not be corrected completely with the presented method, as the discretization effect superimposes with other causes of the Epps effect such as asynchronous \cite{muennix09b} or lagged \cite{reno03,toth09} time series. However, we were able to quantify the contribution of this particular effect to the Epps effect. Our results show, that the discretization effect can be responsible for up to 40\% of the Epps effect, which we define as the difference between the correlation coefficient at a given time and its saturation value. The contribution is particularly large for stocks that are traded at low prices.

\section{Conclusion}
\label{s:conclusion}
We demonstrated the impact of the tick-size on the microstructure of financial returns. This structure can lead to a change in the shape of the distributions of returns and price changes. If a stock exhibits a large price change in the observed period of time, the composition of the return distribution can lead to heavier tails. We also showed that the return distribution consists of return subset distributions that are more sparsely distributed than the complete distribution.

Furthermore, we demonstrated that the discretization effects can distort the calculation of correlation coefficients, especially if the stocks are traded at low prices. We showed that the erroneous correlation coefficient is mainly caused by an improper normalization of the returns. This distortion depends on the impact of the discretization, which grows for small return intervals. Therefore the observed behavior contributes to the Epps effect.

We developed a method to compensate these discretization effects, which we validated in a model setup. The compensation is only based on the tick-size. Despite the interpolation of the price change distribution, the compensation is parameter-free. This method was also applied to market data. We were able to identify and compensate the impact of the tick-size on the correlation coefficient. The results indicate that the discretization error makes a sizable contribution to the Epps effect for stocks that are traded at low prices.

\section*{Acknowledgements}
M.C.M acknowledges financial support from Studienstiftung des deutschen Volkes.
\begin{appendix}
\section{Stock ensembles}
\begin{table}[h]
\caption{Ensemble of 50 stocks from the S\&P 500 index. The stocks provide the highest relation between the mean price $\langle S \rangle$ and its standard deviation $\sigma_{S}/\langle S \rangle$ with at least 1000 trades per day.}
\label{tab:var}
\centering
{\tiny
\begin{tabular}{| l | l | l | r | r | r | r |}
\hline
 Symbol & Name & Stock exchange & $\langle S \rangle$ & $\sigma_{S}$ & $\sigma_{S}/\langle S \rangle$ & Average trades \\
 \hline 
BSC & Bear Stearns Cos. & NYSE & 133.14 & 23.90 & 0.180 & 25529\\
SII & Smith International & NYSE & 57.12 & 10.46 & 0.183 & 13949\\
FRE & Federal Home Loan Mtg. & NYSE & 57.87 & 10.80 & 0.187 & 21389\\
LSI & LSI Corporation & NYSE & 7.93 & 1.49 & 0.187 & 20497\\
PCAR & PACCAR Inc. & NASDAQ & 74.10 & 13.89 & 0.187 & 11376\\
JCP & Penney (J.C.) & NYSE & 69.49 & 13.08 & 0.188 & 15925\\
CVG & Convergys Corp. & NYSE & 21.77 & 4.16 & 0.191 & 5147\\
CBG & CB Richard Ellis Group & NYSE & 31.58 & 6.08 & 0.193 & 13992\\
LIZ & Liz Claiborne  Inc. & NYSE & 35.80 & 6.97 & 0.195 & 6330\\
BC & Brunswick Corp. & NYSE & 28.09 & 5.54 & 0.197 & 5387\\
CNX & CONSOL Energy Inc. & NYSE & 45.20 & 8.99 & 0.199 & 12842\\
AKAM & Akamai Technologies Inc & NASDAQ & 42.89 & 8.63 & 0.201 & 22620\\
MCO & Moody's Corp & NYSE & 56.88 & 11.87 & 0.209 & 16294\\
RSH & RadioShack Corp & NYSE & 24.72 & 5.17 & 0.209 & 14454\\
LUK & Leucadia National Corp. & NYSE & 38.25 & 8.05 & 0.211 & 3429\\
FLR & Fluor Corp. (New) & NYSE & 115.34 & 24.90 & 0.216 & 7413\\
NCC & National City Corp. & NYSE & 30.54 & 6.71 & 0.220 & 17876\\
LXK & Lexmark Int'l Inc & NYSE & 48.47 & 10.91 & 0.225 & 8838\\
DF & Dean Foods & NYSE & 32.98 & 7.49 & 0.227 & 6647\\
MBI & MBIA Inc. & NYSE & 58.90 & 13.50 & 0.229 & 17571\\
FCX & Freeport-McMoran Cp \& Gld & NYSE & 82.68 & 19.07 & 0.231 & 45292\\
FHN & First Horizon National & NYSE & 34.09 & 7.88 & 0.231 & 7319\\
ESRX & Express Scripts & NASDAQ & 70.84 & 16.43 & 0.232 & 11623\\
JNPR & Juniper Networks & NASDAQ & 27.11 & 6.35 & 0.234 & 33006\\
DDS & Dillard Inc. & NYSE & 28.97 & 6.79 & 0.234 & 8291\\
CMI & Cummins  Inc. & NYSE & 117.32 & 47.60 & 0.406 & 9909\\
JNY & Jones Apparel Group & NYSE & 26.09 & 6.24 & 0.239 & 5803\\
MON & Monsanto Co. & NYSE & 70.65 & 16.94 & 0.240 & 17403\\
SOV & Sovereign Bancorp & NYSE & 20.13 & 4.84 & 0.240 & 10837\\
CMCSA & Comcast Corp. & NASDAQ & 27.05 & 6.64 & 0.246 & 55284\\
OMX & OfficeMax Inc. & NYSE & 39.68 & 9.85 & 0.248 & 6009\\
WM & Washington Mutual & NYSE & 36.55 & 9.09 & 0.249 & 39145\\
KG & King Pharmaceuticals & NYSE & 16.29 & 4.15 & 0.254 & 9548\\
JEC & Jacobs Engineering Group & NYSE & 71.33 & 32.20 & 0.451 & 5501\\
CIT & CIT Group & NYSE & 46.43 & 12.31 & 0.265 & 11141\\
THC & Tenet Healthcare Corp. & NYSE & 5.66 & 1.51 & 0.267 & 12112\\
KBH & KB Home & NYSE & 37.21 & 10.34 & 0.278 & 16670\\
GME & GameStop Corp. & NYSE & 47.04 & 23.36 & 0.497 & 9619\\
CTX & Centex Corp. & NYSE & 37.79 & 10.53 & 0.279 & 14328\\
CTSH & Cognizant Technology Solutions & NASDAQ & 72.31 & 20.33 & 0.281 & 13112\\
ODP & Office Depot & NYSE & 28.13 & 7.92 & 0.282 & 14062\\
NOV & National Oilwell Varco  Inc. & NYSE & 88.30 & 30.40 & 0.344 & 19716\\
GILD & Gilead Sciences & NASDAQ & 57.24 & 17.77 & 0.310 & 26240\\
ABK & Ambac Financial Group & NYSE & 70.98 & 23.27 & 0.328 & 16261\\
PHM & Pulte Homes  Inc. & NYSE & 21.82 & 7.34 & 0.336 & 15737\\
LEN & Lennar Corp. & NYSE & 35.41 & 11.91 & 0.336 & 16250\\
MTG & MGIC Investment & NYSE & 46.61 & 17.01 & 0.365 & 14053\\
CC & Circuit City Group & NYSE & 13.65 & 5.00 & 0.366 & 16660\\
ETFC & E*Trade Financial Corp. & NASDAQ & 17.65 & 6.86 & 0.389 & 33380\\
CFC & Countrywide Financial Corp. & NYSE & 28.75 & 11.41 & 0.397 & 65703\\
\hline\end{tabular}

}
\end{table}

\begin{table}[h]
\caption{Ensemble of 50 highest correlation stocks stock pairs from the S\&P 500 index that are averagely traded between \$0.01 and \$10.00. The column $\mathrm{var_{corr}}$ refers to the variance of the correlation of a moving 30-day window.}
\label{tab:high1}
\centering
{\tiny
\begin{tabular}{| l | l | l | l | l | l | l | l | l | c | c |}
\hline
\multicolumn{4}{|c|}{Stock 1} & \multicolumn{4}{|c|}{Stock 2} & $\mathrm{corr}$ & $\mathrm{var_{corr}}$ \\
 Symbol & Name & Stock exchange & Mean Price  & Symbol & Name  & Stock exchange  & Mean Price & & \\
 \hline 
F & Ford Motor & NYSE & 8.16 & Q & Qwest Communications Int & NYSE & 8.63 & 0.11 & 0.02\\
Q & Qwest Communications Int & NYSE & 8.63 & CPWR & Compuware Corp. & NASDAQ & 9.44 & 0.12 & 0.03\\
CPWR & Compuware Corp. & NASDAQ & 9.44 & UIS & Unisys Corp. & NYSE & 7.65 & 0.13 & 0.10\\
CPWR & Compuware Corp. & NASDAQ & 9.44 & THC & Tenet Healthcare Corp. & NYSE & 5.66 & 0.14 & 0.03\\
NOVL & Novell Inc. & NASDAQ & 7.21 & F & Ford Motor & NYSE & 8.16 & 0.14 & 0.02\\
F & Ford Motor & NYSE & 8.16 & LSI & LSI Corporation & NYSE & 7.93 & 0.15 & 0.03\\
Q & Qwest Communications Int & NYSE & 8.63 & THC & Tenet Healthcare Corp. & NYSE & 5.66 & 0.15 & 0.03\\
F & Ford Motor & NYSE & 8.16 & CPWR & Compuware Corp. & NASDAQ & 9.44 & 0.15 & 0.02\\
Q & Qwest Communications Int & NYSE & 8.63 & LSI & LSI Corporation & NYSE & 7.93 & 0.16 & 0.03\\
UIS & Unisys Corp. & NYSE & 7.65 & THC & Tenet Healthcare Corp. & NYSE & 5.66 & 0.17 & 0.05\\
Q & Qwest Communications Int & NYSE & 8.63 & UIS & Unisys Corp. & NYSE & 7.65 & 0.18 & 0.05\\
DYN & Dynegy Inc. & NYSE & 8.77 & THC & Tenet Healthcare Corp. & NYSE & 5.66 & 0.18 & 0.02\\
Q & Qwest Communications Int & NYSE & 8.63 & DYN & Dynegy Inc. & NYSE & 8.77 & 0.19 & 0.03\\
DYN & Dynegy Inc. & NYSE & 8.77 & LSI & LSI Corporation & NYSE & 7.93 & 0.21 & 0.03\\
UIS & Unisys Corp. & NYSE & 7.65 & LSI & LSI Corporation & NYSE & 7.93 & 0.22 & 0.05\\
LSI & LSI Corporation & NYSE & 7.93 & THC & Tenet Healthcare Corp. & NYSE & 5.66 & 0.22 & 0.02\\
NOVL & Novell Inc. & NASDAQ & 7.21 & UIS & Unisys Corp. & NYSE & 7.65 & 0.22 & 0.04\\
CPWR & Compuware Corp. & NASDAQ & 9.44 & LSI & LSI Corporation & NYSE & 7.93 & 0.23 & 0.04\\
F & Ford Motor & NYSE & 8.16 & UIS & Unisys Corp. & NYSE & 7.65 & 0.23 & 0.04\\
DYN & Dynegy Inc. & NYSE & 8.77 & CPWR & Compuware Corp. & NASDAQ & 9.44 & 0.24 & 0.03\\
NOVL & Novell Inc. & NASDAQ & 7.21 & CPWR & Compuware Corp. & NASDAQ & 9.44 & 0.25 & 0.06\\
NOVL & Novell Inc. & NASDAQ & 7.21 & DYN & Dynegy Inc. & NYSE & 8.77 & 0.26 & 0.04\\
NOVL & Novell Inc. & NASDAQ & 7.21 & LSI & LSI Corporation & NYSE & 7.93 & 0.29 & 0.07\\
DYN & Dynegy Inc. & NYSE & 8.77 & UIS & Unisys Corp. & NYSE & 7.65 & 0.29 & 0.03\\
F & Ford Motor & NYSE & 8.16 & DYN & Dynegy Inc. & NYSE & 8.77 & 0.39 & 0.03\\
\hline\end{tabular}

}
\end{table}

\begin{table}[h]
\caption{Ensemble of 50 highest correlation stocks stock pairs from the S\&P 500 index that are averagely traded between \$10.01 and \$20.00. The column $\mathrm{var_{corr}}$ refers to the variance of the correlation of a moving 30-day window.}
\label{tab:high2}
\centering
{\tiny
\begin{tabular}{| l | l | l | l | l | l | l | l | l | c | c |}
\hline
\multicolumn{4}{|c|}{Stock 1} & \multicolumn{4}{|c|}{Stock 2} & $\mathrm{corr}$ & $\mathrm{var_{corr}}$ \\
 Symbol & Name & Stock exchange & Mean Price  & Symbol & Name  & Stock exchange  & Mean Price & & \\
 \hline 
CMS & CMS Energy & NYSE & 17.18 & CZN & Citizens Communications & NYSE & 14.34 & 0.37 & 0.05\\
XRX & Xerox Corp. & NYSE & 17.45 & AW & Allied Waste Industries & NYSE & 12.71 & 0.37 & 0.04\\
MU & Micron Technology & NYSE & 11.38 & TER & Teradyne Inc. & NYSE & 15.07 & 0.37 & 0.04\\
IPG & Interpublic Group & NYSE & 11.23 & HBAN & Huntington Bancshares & NASDAQ & 19.95 & 0.38 & 0.03\\
TE & TECO Energy & NYSE & 16.96 & HBAN & Huntington Bancshares & NASDAQ & 19.95 & 0.38 & 0.05\\
TE & TECO Energy & NYSE & 16.96 & EP & El Paso Corp. & NYSE & 16.08 & 0.38 & 0.02\\
AN & AutoNation  Inc. & NYSE & 19.95 & IPG & Interpublic Group & NYSE & 11.23 & 0.39 & 0.03\\
AW & Allied Waste Industries & NYSE & 12.71 & HBAN & Huntington Bancshares & NASDAQ & 19.95 & 0.39 & 0.03\\
LUV & Southwest Airlines & NYSE & 14.78 & HBAN & Huntington Bancshares & NASDAQ & 19.95 & 0.39 & 0.03\\
DUK & Duke Energy & NYSE & 19.34 & CMS & CMS Energy & NYSE & 17.18 & 0.40 & 0.03\\
JDSU & JDS Uniphase Corp. & NASDAQ & 14.72 & MU & Micron Technology & NYSE & 11.38 & 0.40 & 0.07\\
AN & AutoNation  Inc. & NYSE & 19.95 & TER & Teradyne Inc. & NYSE & 15.07 & 0.40 & 0.02\\
SLE & Sara Lee Corp. & NYSE & 16.73 & EP & El Paso Corp. & NYSE & 16.08 & 0.41 & 0.09\\
CNP & CenterPoint Energy & NYSE & 17.52 & EP & El Paso Corp. & NYSE & 16.08 & 0.41 & 0.03\\
TSN & Tyson Foods & NYSE & 19.06 & ETFC & E*Trade Financial Corp. & NASDAQ & 17.65 & 0.41 & 0.05\\
TER & Teradyne Inc. & NYSE & 15.07 & HBAN & Huntington Bancshares & NASDAQ & 19.95 & 0.42 & 0.04\\
CMS & CMS Energy & NYSE & 17.18 & CNP & CenterPoint Energy & NYSE & 17.52 & 0.43 & 0.06\\
HCBK & Hudson City Bancorp & NASDAQ & 13.85 & HBAN & Huntington Bancshares & NASDAQ & 19.95 & 0.44 & 0.03\\
DUK & Duke Energy & NYSE & 19.34 & WIN & Windstream Corporation & NYSE & 14.26 & 0.44 & 0.11\\
LUV & Southwest Airlines & NYSE & 14.78 & AN & AutoNation  Inc. & NYSE & 19.95 & 0.45 & 0.03\\
CMS & CMS Energy & NYSE & 17.18 & EP & El Paso Corp. & NYSE & 16.08 & 0.45 & 0.03\\
TE & TECO Energy & NYSE & 16.96 & CMS & CMS Energy & NYSE & 17.18 & 0.46 & 0.04\\
TE & TECO Energy & NYSE & 16.96 & DUK & Duke Energy & NYSE & 19.34 & 0.47 & 0.05\\
AN & AutoNation  Inc. & NYSE & 19.95 & HBAN & Huntington Bancshares & NASDAQ & 19.95 & 0.49 & 0.04\\
TE & TECO Energy & NYSE & 16.96 & CNP & CenterPoint Energy & NYSE & 17.52 & 0.51 & 0.04\\
\hline\end{tabular}

}
\end{table}

\end{appendix}
\clearpage

\bibliographystyle{elsarticle-num}
\bibliography{Manuskript_arxiv_v4.bbl}
\end{document}